\documentstyle[prl,aps]{revtex} 
\def\={&=&\mbox{}} 

\def\>{& &\!\!\!\!\!\!\mbox{}}

\def\mk#1{{\mkern#1mu}}
\def\vec#1{{\ifmmode
	\mathchoice{\mbox{\boldmath$\displaystyle#1\mk{-1}$}}
		{\mbox{\boldmath$\textstyle#1\mk{-1}$}}
		{\mbox{\boldmath$\scriptstyle#1\mk{-1}$}}
		{\mbox{\boldmath$\scriptscriptstyle#1\mk{-1}$}}\else
	{\mbox{\boldmath$#1$}}\fi}}

\begin{document}

\draft

\preprint{IPT-98-}

\title
{ Hubbard chain with a Kondo impurity }

\author
{P.-A. Bares and K. Grzegorczyk }

\address
{
Institut de Physique Th\'{e}orique,
\'{E}cole Polytechnique F\'{e}d\'{e}rale de Lausanne,
CH-1015 Lausanne 
}


\maketitle

\begin{abstract}

A Bethe Ansatz solution of a (modified) Hubbard chain with a Kondo impurity of 
arbitrary spin $S$ at a
highly symmetric line of parameter space is proposed and explored. 
Our results confirm the
existence of a strong-coupling (line of) fixed-point(s) with ferromagnetic 
Kondo coupling
as first hypothetized by Furusaki and
Nagaosa on the basis of perturbative renormalization group calculations.
For on-site Hubbard repulsion and
ferromagnetic Kondo exchange, the ground state has spin $S-1/2$,
{\it i.e.} , is a singlet when $S=1/2$.
The contributions of the
impurity to the magnetic susceptibility and low-temperature
specific heat are discussed. 
While the Wilson ratio is unity in the half-filled band, 
it is found to be a 
function of density and interaction away from half-filling. 
\end{abstract}



The recent developments in the field of high-temperature superconductors 
and heavy-fermion materials have stimulated a renewal of interest into the 
theory
of impurities in quantum liquids and of strongly correlated Fermi-systems. 
The search for non-Fermi-liquid fixed points has
gavitated around two prototypal models and their respective descendants:
the Hubbard- ($t-J$, etc.) and the Anderson 
(single impurity and periodic Kondo, etc.) models.
In the last decades, these Hamiltonians and their associated families 
have become the most extensively 
studied many-fermion problems in condensed matter theory. The
progress has been tremendous and the literature in the field is considerable.
For
the sake of
brevity , we 
refer to a few
reviews
\cite{Hubbardreview92}
and the numerous references therein.\\
In the present letter, we want to discuss a Hubbard chain with a Kondo impurity
that is integrable on a particular line of parameter space. Our model emerges 
as a hybride of the one-dimensional Hubbard model \cite{LiebWu} and the standard
Kondo model \cite{Andrei,TsvelikWiegmann}:
\begin{eqnarray}
H\=-t\sum_{i,\sigma}\left( c_{i\sigma}^\dagger c_{i+1\sigma}+\rm{h.c.}\right)+ 
\frac{J}{2}\sum_{\sigma\sigma '}c_{0\sigma}^\dagger
\vec{\sigma}_{\sigma\sigma'}c_{0\sigma'}\cdot{\bf S}\nonumber\\
\>+V\sum_\sigma n_{0\sigma}+U\sum_i n_{i\uparrow}n_{i\downarrow}
+\mu\sum_i\left(n_{i\uparrow}+n_{i\downarrow}\right)-\frac{h}{2}
\sum_{i}\left(n_{i\uparrow}-n_{i\downarrow}\right)-h\gamma S
\label{eq:hamiltonian}
\end{eqnarray}
where $c_{i,\sigma }^{\dagger}$ creates an electron of spin $\sigma=\uparrow, 
\downarrow$
in a Wannier state centered 
at site $i$, $n_{i\sigma}=c_{i,\sigma }^{\dagger}c_{i,\sigma }$;
$t$ denotes the hopping integral, $U$ the screened on-site Coulomb repulsion, 
$J$ 
the exchange coupling the localized impurity spin 
$S$
to the conduction electrons and $V$ a local charge interaction. 
In equation (\ref{eq:hamiltonian}) we 
have
included the chemical potential $\mu$, the Zeeman-term with 
$h=g_e \mu_B H$ and $\gamma=g_s/g_e$ the gyromagnetic ratio for the ion impurity 
in units of the conduction-band electron $g_e$ factor. We assume a 
chain of $L$ sites and 
impose periodic boundary conditions.
The model of equation 
(\ref{eq:hamiltonian}) ignores important physical ingredients \cite{Nozieres} 
such as
orbital degeneracy, spin orbit coupling,
crystal-field effects, and so on.\\
To our knowledge, the Hubbard-Kondo (H-K) model of equation 
(\ref{eq:hamiltonian})
has not been solved previously, though various solutions of
related models on the
continuum and lattice have been proposed 
(see\cite{tJBares,WangVoit,Schlottmann,YQLiBares,Schlottmann98}). 
As is well known the Kondo model can be derived from the Anderson model 
(see \cite{Schrieffer-Wolff}) as a low-energy effective hamiltonian via 
a Schrieffer-Wolff transformation in the regime where the
Anderson model has a local moment and antiferromagnetic exchange with the
conduction electrons (a potential scattering also arises in the reduction 
process).
We note that for the H-K model considered in this work we assume {\it a priori\ 
}
the magnitude of the exchange $J$ to be arbitrary. Furthermore, due to the
presence of the electron-electron interactions, the H-K model is 
a genuine
one-dimensional model in contrast to the original three-dimensional Kondo model,
which is solved exactly by keeping only s-wave states, {\it i.e.}, 
left-going electrons
scattered forward by the Kondo exchange operator.  
The Hamiltonian (\ref{eq:hamiltonian}) inherits various
symmetries of its progenitors (Hubbard and Kondo)
provided appropriate rotations in spin space are performed \cite{BaresGrzeg}.
The most remarkable property of the model (when appropriately modified, see
below)
is however its integrability at $J=-U/2$ 
and 
$V={U\over 2}S$ or  $V=-{U\over 2}(S+1)$, as we now discuss.\\
To solve the eigenvalue problem associated with the Hamiltonian of equation 
(\ref{eq:hamiltonian}), 
we
make the following Ansatz for the N-fermion problem:
\begin{equation}
\Psi_{\sigma_1\cdots\sigma_N}(x_1,\ldots,x_N)=\sum_{Q,P}
\Theta_0(x_{\widetilde{Q}})\Theta(x_Q)(-1)^P 
A_{\sigma_{Q1}\cdots\sigma_{QN}}({\widetilde Q}|PQ)\exp
\left(i\sum k_{P_j}x_j\right)
\label{eq:Ansatz}
\end{equation}
where $\Theta({x}_Q)=\Theta(x_{Q1}<...<x_{QN})=1$ 
when the ordering associated to the permutation
$Q$ of the coordinates is respected and 
otherwise zero; 
$\Theta_0({x}_{\widetilde Q})=\prod_j^N\Theta(\nu_j x_j)$, where
$\nu_j({\widetilde Q})=\pm 1$ according to whether the coordinate is to the 
left
or to the right of the impurity as specified by the permutation 
${\widetilde Q}$ 
of $(N+1)$ objects, and where $\Theta(x)=1$ if $x>0$, and otherwise zero with 
the prescription
$\Theta(0)=1/2$.\\
A straightforward substitution of the Ansatz (\ref{eq:Ansatz}) 
into the Schr\"odinger equation 
$H\Psi=E\Psi$
leads to difficulties which however are only apparent. Indeed, 
we need a prescription of how to
treat the electrons as they approach the impurity. For the sake of simplicity
consider two particles away from the impurity with amplitude
$\Psi_{\sigma_1,\sigma_2}(x_1,x_2)$. The Hamiltonian is modified near the
impurity as follows:
when, particle one, say, hops from the right onto 
the impurity, the amplitude becomes $\Psi_{\sigma_1,\sigma_2}(0^{+},x_2)$, while
if it hops from the left we obtain $\Psi_{\sigma_1,\sigma_2}(0^{-},x_2)$. Here
$0^{\pm}$ means that we have taken the limit to zero from the right (left) when
approaching the impurity from the right (left).
The amplitude for particle one to be at site $x_1=0$ (impurity site) is defined
by $\Psi_{\sigma_1,\sigma_2}(0,x_2)=[\Psi_{\sigma_1,\sigma_2}(0^{+},x_2)+
\Psi_{\sigma_1,\sigma_2}(0^{-},x_2)]/2$, and so there is no ambiguity as to a
particle sitting on the impurity site. Similarly for particle two.
Furthermore, due to the presence of the impurity, the
four amplitudes $\Psi_{\sigma_1,\sigma_2}(0^{\pm},0^{\pm})$ cannot 
coincide for our
ansatz to be consistent. Yet, we can define the amplitude for two particles of 
opposite
spin to be at the
impurity site as the average of these four amplitudes. With these definitions,
the eigenvalue problem can be dealt with in the usual way near the impurity 
while at
the impurity site additional terms occur, that 
cancel exactly those coming from the electron-impurity interaction.  
The idea is readily
extended to the $N$-particle case \cite{BaresGrzeg}. 
That the low-energy long-wave length physics of the original Hubbard model 
with Kondo impurity 
should not be affected by our strategy can be seen as follows: the model we
discuss here can be viewed as a lattice regularization of
a continuum 
Hamiltonian solved in Ref.\cite{YQLiBares} (see comments in the bibliography).
We will provide a  complete discussion of this 
point together with numerical computations elsewhere~\cite{BaresGrzeg}.\\
The Schr\"odinger equation imposes
linear relations between the coefficients $A ({\widetilde Q}|PQ)$ 
in the Ansatz (\ref{eq:Ansatz}) and determines
the many-body scattering matrix. The electron-electron 
scattering
matrix is identical to that found in the Hubbard model (see
\cite{LiebWu}) while
the electron-impurity scattering matrix reads 
\begin{equation}
R_{js}(\sin k_j)=\frac{2 i t\sin k_j-({\widetilde V}+JP^{js})}
{2 i t\sin k_j +({\widetilde V}+JP^{js})} \quad ,
\label{eq:eiscattering}
\end{equation}
where $\sin k_j$ is the rapidity of the colliding
electron, $P^{js}=\frac{1}{2}+\vec{\sigma}\cdot {\bf S}$ and 
${\widetilde V}=V-J/2$.
The many-body S-matrix factorizes into two-body scattering matrices
provided the consistency conditions (Yang-Baxter
equations) are fulfilled. Since the latter are identical in form
to those occuring in the Kondo model, we refer the reader to the literature
(see for example \cite{Andrei,TsvelikWiegmann}).
The Yang-Baxter equations require then $ J=-{U\over 2}$ and ${\widetilde V}=\pm 
{U\over 2}(S+1/2)$.\\
We can express the periodic boundary conditions in terms of the two-body
scattering matrices, 
$T_jA_{\sigma_1\cdots\sigma_NS}(I|I)=e^{ik_jL}A_{\sigma_1\cdots\sigma_NS}(I|I)$, 
where $I$ denotes the identity permutation and 
$T_j=S_{jj+1}\ldots S_{jN}R_{js}S_{j1}\ldots S_{jj-1}$
represents the transfer matrix that carries a particle around the ring.
The eigenvalue problem can now be solved by the technique of the
monodromy matrix (Quantum Scattering Method, see for example
\cite{TsvelikWiegmann,KorepinBI}). For ${\widetilde V}=\pm{U\over 2}(S+1/2)$,
we find the
following set of coupled algebraic equations :
\begin{mathletters}
\label{eq:BAequ}
\begin{equation}
 e^{ik_jL}=\frac{\alpha_j\pm ig(S+\frac{1}{2})}
{\alpha_j+ig(S+\frac{1}{2})}\prod_{\beta=1}^{N_s}
\frac{\alpha_j-\Lambda_\beta+i\frac{g}{2}}
{\alpha_j-\Lambda_\beta-i\frac{g}{2}}
\label{equationa}
\end{equation}
\begin{equation}
\mbox{}\ \ -\prod_{\beta=1}^{N_s}
\frac{\Lambda_\beta-\Lambda_\alpha+ig}
{\Lambda_\beta-\Lambda_\alpha-ig}=
\frac{\Lambda_\alpha-igS}
{\Lambda_\alpha+igS}\prod_{j=1}^{N_c}
\frac{\alpha_j-\Lambda_\alpha +i\frac{g}{2}}
{\alpha_j-\Lambda_\alpha -i\frac{g}{2}}\label{equationb}
\end{equation}\end{mathletters}
where $N_c=N$ is the number of charges and $N_s$ the number 
of down spins in the system, $\alpha_j=\sin k_j$ ($j=1\ldots N_c$), 
denote the charge rapidities, 
$\Lambda_{\alpha}$ ($\alpha=1\ldots N_s$) the spin rapidities 
and $g=U/2t$. 
Equations (\ref{eq:BAequ}) resemble those derived by Lieb and Wu for 
the Hubbard model, yet
include an additional phase shift due to the presence of the impurity.
Taking the logarithm of equations (\ref{eq:BAequ}) leads to a system of coupled 
transcendental equations parametrized by two sets of quantum numbers 
$I_j^{\alpha}$ 
$(\alpha =c, s)$, which 
are intergers or half-odd
integers depending on the values of $N_c$, $N_s$ and $L$. Notice that the
impurity shifts the quantum numbers of the Hubbard model and 
that in contrast to
the Kondo model, the electron-impurity phase-shift depends on the state of
motion of the charges and spin degrees of freedom through their rapidities. 
In the thermodynamic limit, we 
introduce the distributions $\rho_c(k)$ and $\rho_s(\lambda)$
of the "charge" and "spin" rapidities  \cite{LiebWu} that obey coupled integral
equations of Fredholm type. Since the method is by now standard, we refer the
reader for the details to \cite{BaresGrzeg} and 
discuss here the physics of the impurity.\\
In this letter we focus on the contribution of the impurity to the magnetization 
and the low-temperature thermodynamic properties of the system.
The magnetization in zero-field $h=0$ 
is given by $M=M_h+{1\over L}M_i$ with $M_h=0$ and $M_i=S-1/2$, 
{\it i.e.}, the
ground-state is $2S$-fold degenerate and the spin of the impurity is partially
screened. When $S=1/2$ the ground-state is a \it singlet \rm despite the
ferromagnetic Kondo exchange coupling. 
This result is somewhat counter-intuitive, especially in
view of a theorem due to Mattis \cite{Mattis} which states that 
for a ferromagnetic Kondo model in the absence of electron-electron interactions  
the ground-state has spin $S+1/2$, {\it i.e.}, is a triplet for spin $S=1/2$. Our 
results
can however be understood
 if we remember that as soon as $U$ is finite, we scale to a
(line of) strong-coupling fixed point(s): the screening of the local spin is 
a consequence of the presence of relevant (with respect to the Luttinger liquid
fixed point) local backscattering at the impurity site 
as seen from bosonization of the H-K model \cite{RenAnderson,FurusakiNagaosa}. 
The existence of a
strong-coupling 
fixed-point has tentatively been suggested by Furusaki and Nagaosa 
\cite{FurusakiNagaosa} using perturbative renormalization group 
techniques. The ferromagnetic Kondo exchange coupling makes the electron spin on
the impurity ion to
align with the localized magnetic moment while the neighboring 
electrons rearrange in such
a way as to screen the
resulting
enhanced local magnetization.\\ 
At arbitrary magnetic fields, the problem has to be solved numerically 
\cite{BaresGrzeg}. 
In the following, we consider 
the weak magnetic field regime which can be treated analytically.
Let us first discuss the half-filled band 
$n_c=1$. Because umklapp-scattering 
is relevant at half-filling, this
case was not discussed in ref. \cite{FurusakiNagaosa}.
In the presence of a weak $h>0$, the bulk magnetization
is (see for example \cite{Shiba,FrahmKorepin})  
$M_h={h\over 4\pi t} {I_0(\pi/g)\over I_1(\pi/g)}$, where 
$I_{\nu}(z)$, for $\nu =0,1$,
denote the 
Bessel functions of the first kind of imaginary argument, 
and the
contribution of the impurity is obtained by the Wiener-Hopf method \cite{Morse}
as 
\begin{eqnarray}
M_i\approx S-\frac{1}{2}+\frac{1}{\pi^{3/2}}\int_0^\infty
\frac{d\omega}{\omega}\sin\left[(2S-1)\pi\omega\right]
\left(\frac{\omega}{e}\right)^{\omega}\Gamma
\left(\frac{1}{2}-\omega\right)\exp\left({2\omega\ln(h/T_H)}\right) \nonumber\\
 \mbox{}\hspace{0.5cm} +\frac{1}{\sqrt{\pi}}\sum_{k=0}^\infty
\frac{(-1)^k}{k!(k+1/2)}\left(\frac{k+1/2}{e}\right)^{(k+1/2)}
\left(\frac{h}{T_H}\right)^{(2k+1)}\cos[2\pi(S-1/2)(k+1/2)]
\label{eq: half-filled}
\end{eqnarray}
where $g|\ln(h/T_H)|\gg 1$ and where we have set $\gamma=1$.
Note that formula (\ref{eq: half-filled}) resembles the one found in the
antiferromagnetic Kondo model.
Moreover, in equation (\ref{eq: half-filled}) the coupling constant $g$ 
enters through the expression $\ln(h/T_H)$.
The magnetic scale $T_H$ (in units of $k_B$ Bolzmann) has been  defined by
$T_H=(2\pi/e)^{1/2} T_K$, where  ``the Kondo-Hubbard temperature'', 
$T_K=4t I_1(\pi/g)$, 
represents a strong coupling scale 
controlled by the band-width 
of the conduction-band
and a Bessel function that interpolates betwen the strong- 
($g\rightarrow \infty$) and weak- ($g\rightarrow 0$) coupling limits of
the Hubbard model. We expect that for $T\ll T_K$
all physical
quantities depend only on the energy scale $T_K$.
The case ($U=-2J<0$) of attractive on-site interaction
and antiferromagnetic exchange 
is also interesting and will be discussed in \cite{BaresGrzeg}.\\
We distinguish the compensated ($S=1/2$)
and uncompensated ($S>1/2$) cases. 
When $S=1/2$, the magnetization can be written as
\begin{equation}
M_i\approx\frac{1}{\sqrt{\pi}}\sum_{k=0}^\infty
\frac{(-1)^k}{k!(k+1/2)}\left(\frac{k+1/2}{e}\right)^{(k+1/2)}
\left(\frac{h}{T_H}\right)^{(2k+1)}
\label{eq: hfmagnet1}
\end{equation}
where $h\ll T_H$. To lowest order in the field the magnetization is linear in 
$h$, 
{\it i.e.}, the impurity magnetic susceptibility is a constant  
$\chi_i\approx {1/\pi T_K}$. This is consistent with the picture that 
the impurity spin is screened at zero
temperature by the conduction
spin carriers. Here, $T_K$ represents the ``binding energy'' of a complex of
electrons and the impurity spin.
The strong-coupling limit of the Hubbard model $g\gg 1$ yields
an enhanced susceptibility $\chi_i\approx g/2\pi^2 t$ identical 
to the bulk
susceptibility \cite{FrahmKorepin}
per site $\chi_h\approx {U/ 4\pi^2 t^2}$, {\it i.e.}, 
at weak but finite field the electrons are less and less effective in screening
the impurity spin with increasing $g=U/2t=-J/2t\rightarrow \infty$. 
The zero-temperature
susceptibility diverges when $g\rightarrow \infty$
as expected for decoupled spins. 
Let us recall that in this limit the model maps 
(via second order virtual processes)
to a Heisenberg chain with an 
antiferromagnetic effective exchange $J_e\approx 4t^2/U$ and an electron
spin trapped at the impurity site.
A simple picture is that of a complex of four spins built from a
doublet of the impurity spin and two neighboring electron spins 
coupled via antiferromagnetic effective interaction to an
electron spin sitting on the impurity site.
In the limit of weak coupling, $g\ll 1$, the magnetic susceptibility behaves as
$\chi_i\approx { 1\over 2t\sqrt{2 g }}\exp\left(-\frac{\pi}{g}\right)$ 
while the bulk susceptibility $\chi_h\approx (2\pi t)^{-1}$ is finite, {\it i.e.},
the impurity susceptibility is strongly suppressed. Here, the naive 
(mean-field) picture of a spin-density wave coupled to a local magnetization
applies: the former has effectively swallowed the impurity spin.
While in a singlet ground-state,
we can pass continuously, by varying the Hubbard repulsion (Kondo
exchange)
$U=-2J$, from a regime where the impurity spin has
disappeared locally to a regime 
where the impurity site has decoupled from the remaining
spins.\\
For $S>1/2$ the dominant terms in the
impurity magnetization reflect the physics of a strong coupling
regime,
\begin{equation}
M_i\approx\left( S-1/2\right)
\left[ 1+\frac{1}{\ln(T_H/h)}
-\frac{\ln(\ln(T_H/h))}{2\left[\ln(T_H/h)\right]^2}\right]
\label{eq: magS}
\end{equation}
The corrections to the above formula are exponentially small.
The resulting susceptibility is singular in the limit of small field, a result
that can be interpreted in terms of a reduced spin $(S-1/2)$ having a
ferromagnetic interaction with the conduction-band electrons.\\
It can be shown \cite{BaresGrzeg} that at half-filling
(as in the standard antiferromagnetic Kondo model)
there is a duality on substitution of $S$ by $(S-1/2)$
between the low ($T\ll T_K$) and  high ($T\gg T_K$)
 temperatures as seen for example in the magnetic 
susceptibility  
\begin{equation}
\chi_i(T)=
\left\{\begin{array}{ll}
\displaystyle{\frac{1}{3}(S^2-1/4)\left[1-\frac{1}{\ln(T/T_K)}+
\frac{\ln|T/T_K|}{\ln^2(T/T_K)}+\dots\right]} & \text{for $T\ll T_K$}, 
\\[1.2em]
\displaystyle{\frac{1}{3}S(S-1)\left[1-\frac{1}{\ln(T/T_K)}+ 
\frac{\ln|T/T_K|}{\ln^2(T/T_K)}+\dots\right]}
&  \text{for $T\gg T_K$}.
\end{array}
\right.
\end{equation}
The low temperature specific heat at half-filling for spin $S=1/2$
has contributions only from the
spin degrees of freedom, {\it i.e.},
$C_v=C_v^h+C_v^i/L={\pi T\over 3 v_s}
\left[ 1+\rho_s^i(q_s)/L
\rho_s^h(q_s)\right]$, where the spinon velocity $v_s=2t 
{I_1(\pi/g)/I_0(\pi/g)}$ 
and the bulk specific heat $C_v^h={I_0(\pi/g)/2\pi tI_1(\pi/g)}$ 
(see see for example \cite{Takahashi}).
 For $S=1/2$, we have $C_v^i/C_v^h=1/I_0(\pi/g)$,
{\it i.e.}, $C_v^i/C_v^h(g\rightarrow\infty)\approx 1$ and 
$C_v^i/C_v^h(g\rightarrow 0)\approx \pi \sqrt{2g}
\exp{\left(-\frac{\pi}{g}\right)}$, 
again a manifestation of the fact that the itinerant electrons become more
effective in screening the impurity in the limit  
$g\rightarrow 0$.  
At half-filling and for spin $S=1/2$, 
the Wilson ratio is therefore : $R^{(1/2)}={\chi_i/\chi_h\over C_v^i/C_v^h}=1$. 
Near the saturation field $h_c$ (see for example \cite{FrahmKorepin}), the
susceptibility and specific heat behave as 
$\chi_i/ \chi_h=C_v^i/ C_v^h\approx 2/ S g F(g)$ where
$F(g)$ is a function of $g$.
We thus
infer $R^{(S>1/2)}=1$, for $h$ near $h_c$, as in zero-field. 
In the weak coupling limit, we have 
$\chi_i/\chi_h=C_v^i/ C_v^h\approx 1/ gS$ while
in the strong coupling limit, the above ratio becomes independent 
of the coupling
constant $g$ , {\it i.e.},
$\chi_i/\chi_h=C_v^i/ C_v^h\approx 1/2S$.\\
We now consider the strong coupling limit at arbitrary band filling.
For a compensated impurity, the magnetization at weak fields is 
$M_i\approx 2h/\pi h_c$,
which is proportional to the bulk magnetization so that $\chi_i/\chi_h\approx
1/n_c$. The specific heat ratio simplifies in this limit and 
the Wilson ratio becomes 
\begin{equation} 
R^{(1/2)}\approx 1+
\frac{\pi}{2 g \sin(\pi n_c)}
\left( 1-\frac{\sin(2\pi n_c)}{2\pi n_c }\right)+O(g^{-2})\;\; ,
\label{eq: Wilsondemi} 
\end{equation}
ergo $R^{(1/2)}$ is not a universal number, {\it i.e.}, is a function of the 
interaction $g$
and density. The above expression is valid provided $\delta=1-n_c\gg 1/g $.\\
At small density $n_c\ll 1$ we can  estimate the specific heat 
and susceptibility ratios for $S=1/2$. We find a Wilson ratio 
$R^{(1/2)}\approx 1+\pi^2 n_c/g$, again a non-universal number as in the
large $g$ limit.
For spin $S>1/2$ and away from half-filling, the magnetization is identical in
form to that found at half-filling on the substitution 
$T_H\rightarrow T_H^n$, where the new energy scale reads $T_H^n\approx
(2\pi/e)^{1/2}(2t/g) (\pi n_c)^3$. The physical interpretation of this result 
is similar to
that of the above case, except that now the band width of the conduction
electrons scales to zero with the charge density.\\
In conclusion, we have solved and discussed 
a Hubbard chain with repulsive electron-electron interactions and ferromagnetic
impurity exchange. Our solution confirms the existence of a (line of) 
strong-coupling fixed point(s), which had been tentatively suggested 
by Furusaki and Nagaosa on the basis of
perturbative renormalization.


The authors are grateful to J.-Ph. Ansermet,
C. Gruber, P. Nozi\`eres and T.M. Rice for discussions.

\end{document}